\documentclass[12pt,a4paper]{article}
\usepackage[utf8]{inputenc}
\usepackage[T2A]{fontenc}
\usepackage{amsmath, amssymb}
\usepackage{graphicx}
\usepackage{cite}
\usepackage{hyperref}

\usepackage{etoolbox}
\AtBeginEnvironment{tabular}{%
    \footnotesize\setlength{\tabcolsep}{3.5pt}%
}

\title{Electron-laser vacuum breakdown in head-on collision of relativistic electrons with intense laser pulse}
\author{P.A. Golovinski\\ Voronezh State University, Russia}
\date{}

\begin{document}

\maketitle

\begin{abstract}
The phenomenon of electron-laser vacuum breakdown is the multiple cascade production of electron-positron pairs in head-on collision of a beam of relativistic electrons with an intense laser pulse. This effect was first predicted by the author in 1996 [1] and further developed in [2]. In the present paper, an analytical expression for the total number of produced particles is obtained using the generalized Heitler model. The model results are shown to be in good agreement with the estimates of the pioneering works. An analysis of modern laser facilities (ELI, XCELS, European XFEL, Russian projects) is carried out and estimates of the expected effects are given. At ELI and XCELS class facilities, the quantum nonlinearity parameter can reach 60--150, corresponding to the deeply nonlinear QED regime with multiplicity up to 100 particles per seed electron. Experimental confirmation of the effect is expected in the coming years.
\end{abstract}

\textbf{Keywords:} electron-laser vacuum breakdown, cascade pair production, nonlinear QED, intense laser fields, generalized Heitler model.

\section{Introduction}

The problem of multiple production of electron-positron pairs in strong electromagnetic fields has a long history. Classical works by Schwinger, Breit--Wheeler, and later studies by Nikishov and Ritus laid the foundations of the theory. However, for a long time the question of cascade multiplication of particles, where the produced pairs themselves become sources of new ones, remained open.

In the author's works [1,2], the process called electron-laser vacuum breakdown was theoretically investigated for the first time. In a head-on collision of an ultrarelativistic electron with an intense laser pulse, in the rest frame of the electron the laser photons acquire, due to the Doppler shift, energy sufficient for pair production. The produced electrons and positrons also interact with the field and generate new pairs, leading to the development of a cascade. The process fades as the particle energy decreases below the pair production threshold.

Over the past years, laser facilities of extreme power have been constructed: ELI [3], XCELS [4], European XFEL [5], and projects at IAP RAS and MEPhI are being implemented [6,7]. These facilities allow approaching the regime predicted in [1,2]. Recently, works have appeared that confirm and develop the original ideas [6--8].

The purpose of the present work is to give an analytical description of electron-laser vacuum breakdown based on the generalized Heitler model, to compare with the results of [1,2], and to provide estimates for modern facilities. The main results presented in early publications are also available in the monograph [9] (Chapter 5).

\section{Theoretical model}

Consider the collision of an ultrarelativistic electron with energy \(\varepsilon_0 = \gamma mc^2\) and a laser pulse of intensity \(I\) and frequency \(\omega_0\) (wavelength \(\lambda = 2\pi c/\omega_0\)). In the laboratory frame, the dimensionless field amplitude is
\begin{equation}
a_0 = \frac{eE}{m\omega_0 c} = \sqrt{\frac{I}{I_r}},\qquad 
I_r = \frac{m^2c^3\omega_0^2}{4\pi e^2}\approx 1.37\times10^{18}\left(\frac{1\ \mu\text{m}}{\lambda}\right)^2 \text{ W/cm}^2.
\label{eq:I_r}
\end{equation}

Upon transition to the electron rest frame, the laser photons have energy \(\varepsilon_\gamma' \approx 2\gamma \hbar\omega_0\) due to the Doppler shift. The invariant quantum nonlinearity parameter, which determines the probabilities of processes, is
\begin{equation}
\chi_e = \frac{2\gamma \hbar\omega_0}{mc^2}\,a_0 = 2\gamma\frac{a_0}{a_S},\qquad a_S = \frac{mc^2}{\hbar\omega_0}\approx \frac{0.511\text{ MeV}}{\hbar\omega_0}.
\label{eq:chi}
\end{equation}
For \(\chi_e \sim 1\), pair production processes become significant; for \(\chi_e \gg 1\), the deeply nonlinear QED regime occurs.

According to the idea of [1,2], in the electron rest frame the laser photons produce pairs, and the products are also capable of emitting and producing new pairs. A cascade develops, the number of particles grows, and their energy decreases.

To describe the cascade evolution, we introduce distribution functions for electrons (positrons) \(f_l(\varepsilon,t)\) and photons \(f_\gamma(\varepsilon,t)\) in energy \(\varepsilon\); \(t\) is the development depth. Neglecting spatial inhomogeneity and the back-reaction of particles on the field, the kinetic equations are [8,10]:
\begin{equation}
\begin{aligned}
\frac{\partial f_l}{\partial t} &= \int_{\varepsilon}^{\infty} f_l(\varepsilon',t) w_{e\to\gamma}(\varepsilon',\varepsilon'-\varepsilon)\,d\varepsilon' 
+ 2\int_{\varepsilon}^{\infty} f_\gamma(\varepsilon',t) w_{\gamma\to e}(\varepsilon',\varepsilon)\,d\varepsilon' \\
&\quad - \int_0^{\varepsilon} f_l(\varepsilon,t) w_{e\to\gamma}(\varepsilon,\varepsilon')\,d\varepsilon',\\
\frac{\partial f_\gamma}{\partial t} &= \int_{\varepsilon}^{\infty} f_l(\varepsilon',t) w_{e\to\gamma}(\varepsilon',\varepsilon)\,d\varepsilon' 
- \int_0^{\varepsilon} f_\gamma(\varepsilon,t) w_{\gamma\to e}(\varepsilon,\varepsilon')\,d\varepsilon'.
\end{aligned}
\label{eq:kinetic}
\end{equation}
Here \(w_{e\to\gamma}\) and \(w_{\gamma\to e}\) are the differential probabilities of emission and pair production. The initial condition is \(f_l(\varepsilon,0)=\delta(\varepsilon-\varepsilon_0)\), \(f_\gamma(\varepsilon,0)=0\).

For an analytical solution, we use the generalized Heitler model [11], where processes are considered discrete: an electron with energy \(\varepsilon\) after traversing a length \(L_e\) emits a photon of energy \(k\varepsilon\);
a photon with energy \(\varepsilon\) after traversing a length \(L_\gamma\) produces a pair, with the electron and positron each receiving \(\varepsilon/2\).

Then
\begin{equation}
w_{e\to\gamma}(\varepsilon',\varepsilon) = \frac{1}{L_e}\delta(\varepsilon - k\varepsilon'),\qquad 
w_{\gamma\to e}(\varepsilon',\varepsilon) = \frac{1}{L_\gamma}\delta\!\left(\varepsilon - \frac{\varepsilon'}{2}\right).
\label{eq:heilter}
\end{equation}

Substituting into (2) and integrating over energy yields a system for the total numbers \(N_l(t)=\int f_l d\varepsilon\), \(N_\gamma(t)=\int f_\gamma d\varepsilon\):
\begin{equation}
\frac{dN_l}{dt} = \frac{2}{L_\gamma}N_\gamma,\qquad 
\frac{dN_\gamma}{dt} = \frac{1}{L_e}N_l - \frac{1}{L_\gamma}N_\gamma.
\label{eq:ntot_system}
\end{equation}

The parameter \(k\) has dropped out, i.e., the total number of particles does not depend on the details of the emission spectrum [10]. Setting \(L_e=L_\gamma=L\) (symmetric case), we obtain the solution:
\begin{equation}
N_l(t) = \frac{2}{3}e^{t/L} + \frac{1}{3}e^{-2t/L},\quad 
N_\gamma(t) = \frac{1}{3}\left(e^{t/L}-e^{-2t/L}\right).
\label{eq:ntot_sol}
\end{equation}
The total number of particles is \(N_{\text{tot}}=N_l+N_\gamma = e^{t/L}\).

The cascade stops when the particle energy falls below the pair production threshold \(\varepsilon_{\text{cr}}\). The number of generations \(n = t/L\) is related to the energy by \(\varepsilon \approx \varepsilon_0 2^{-n}\). From \(\varepsilon(n_m)=\varepsilon_{\text{cr}}\) we have \(n_m = \log_2(\varepsilon_0/\varepsilon_{\text{cr}})\). Then the maximum number of particles is
\begin{equation}
N_{\text{tot}}^{\max} \approx \frac{2}{3}\frac{\varepsilon_0}{\varepsilon_{\text{cr}}}.
\label{eq:nmax_final}
\end{equation}

The critical energy is determined from the condition \(\chi(\varepsilon_{\text{cr}})\approx1\). From (1) it follows that
\begin{equation}
\varepsilon_{\text{cr}} \approx \frac{mc^2}{2}\frac{a_S}{a_0}.
\label{eq:ecr}
\end{equation}
For the parameters of [2] (\(\varepsilon_0=800\) GeV, KrF laser, \(I=10^{20}\) W/cm²), we calculate: \(a_0\approx2.12\), \(a_S\approx1.02\times10^5\), \(\chi_e\approx65\), \(\varepsilon_{\text{cr}}\approx12.3\) GeV, \(N_{\text{tot}}^{\max}\approx43\). Reference [2] obtained about 60 particles per electron. Given the approximate nature of the model, the agreement is satisfactory. With \(\chi_{\text{cr}}=0.5\) we get \(N_{\text{tot}}\approx87\), which is closer to the estimate of [2]. Thus, the Heitler model confirms the main conclusions of [1,2].

\section{Estimates for modern facilities}

Table~1 presents the parameters of modern laser facilities. Table~2 shows the calculated values.

\begin{table}[h]
\centering
\caption{Parameters of modern facilities}
\label{tab:params}
\begin{tabular}{|l|c|c|c|c|l|}
\hline
Facility & \(I\), W/cm² & \(\lambda\), nm & \(\tau\), fs & \(\varepsilon_0\), GeV & Reference \\
\hline
Works [1,2] & \(10^{20}\) & 248 & 300 & 800 & [1,2] \\
ELI-NP & \(10^{23}\) & 800 & 20 & 20 & [3] \\
XCELS  & \(10^{24}\) & 910 & 15 & 15 & [4] \\
European XFEL & – & 0.05–0.4 & 20 & 17.5 & [5] \\
IAP/MEPhI projects & \(10^{22}\) & 800–1000 & 30 & 10 & [6,7] \\
\hline
\end{tabular}
\end{table}

\begin{table}[h]
\centering
\caption{Calculated parameters and expected effects}
\label{tab:results}
\begin{tabular}{|l|c|c|c|c|c|}
\hline
Parameter & [1,2] & ELI-NP & XCELS & XFEL & Russian projects \\
\hline
\(a_0\) & 2.12 & 300 & 950 & – & 100 \\
\(a_S\) & \(1.02\cdot10^5\) & \(3.3\cdot10^5\) & \(3.75\cdot10^5\) & \(10^7\)–\(10^8\) & \(3.3\cdot10^5\) \\
\(\chi_e\) & 65 & 71 & 149 & 0.2–2.0 & 10 \\
QED regime & Nonlinear & Deeply nonlinear & Deeply nonlinear & Transitional & Nonlinear \\
\(\varepsilon_{\text{cr}}\), GeV & 12.3 & 0.28 & 0.10 & 10–100 & 0.85 \\
\(N_{\text{max}}\) & 43 & 48 & 100 & 0.1–1.2 & 7.8 \\
 \(n_m\) & 6 & 6 & 7 & 0–4 & 3–4 \\
Energy, MeV & 170 & 0.5–5 & 0.2–2 & 100–1000 & 1–10 \\
Burst duration, fs & 300 & 20 & 15 & 20 & 30 \\
Spatial scale, $\mu$m & 90 & 6 & 4.5 & 6 & 9 \\
\hline
\end{tabular}
\end{table}

For ELI-NP, \(\chi_e\approx71\), up to 50 particles per electron are expected; characteristic energy 0.5–5 MeV. For XCELS, \(\chi_e\approx149\), multiplicity reaches 100, energy 0.2–2 MeV. European XFEL operates in the X-ray range; here \(\chi_e\sim0.2\)–2, multiplicity is small, but the process occurs at high energies. Russian projects give \(\chi_e\approx10\) and multiplicity about 8 particles, which is convenient for method development.

Table~3 presents experimental data on pair production. The closest result was obtained at Astra Gemini [16], where the yield reached \(\sim1\) particle per electron. Reports of "7 particles" usually refer to the total number of positrons, not the yield per electron; proper recalculation gives values many orders of magnitude lower.

\begin{table}[h]
\centering
\caption{Experimental data on pair production}
\label{tab:exper}
\begin{tabular}{|l|c|c|c|c|c|}
\hline
Experiment, year & \(\varepsilon_0\), GeV & \(I\), W/cm² & \(\chi_e\) & Yield per \(e^-\) (exp.) & Reference \\
\hline
SLAC E-144, 1997 & 46.6 & \(1.3\times10^{18}\) & 0.3–0.4 & \(\sim 2\times10^{-4}\) & [12,13] \\
LUX, 2009–2012 & 10 & \(5\times10^{21}\) & 0.8–1.2 & 0.1–0.3 & [14] \\
VULCAN, 2015 & 0.75 & \(5\times10^{20}\) & 0.1–0.2 & \(\sim 10^{-8}\) & [15] \\
Astra Gemini, 2018–2019 & 2 & \(10^{21}\) & 0.5–0.8 & 0.8–1.2 & [16] \\
BELLA, 2021–2023 & 10 & \(10^{20}\) & 0.4–0.6 & 0.5 & [17] \\
\hline
\end{tabular}
\end{table}

Cascade development requires time. ... (остальной текст раздела 3 и Заключения без изменений)

\section{Conclusion}

The theory of electron-laser vacuum breakdown has been developed on the basis of the generalized Heitler model. The analytical expression obtained for the number of produced particles agrees with the pioneering estimates [1,2]. An analysis of modern facilities has been carried out: at ELI and XCELS, multiplicities up to 100 particles per electron are expected. Russian projects yield about 8 particles. It is shown that the duration of modern femtosecond pulses is sufficient for full cascade development. Experimental data have not yet reached the cascade regime. Its realization is expected in the coming years. Further development of the theory should take into account collective effects and include Monte Carlo simulations.


\begin{thebibliography}{99}

\bibitem{Golovinsky1996} Golovinsky P.A. Strong laser-field interaction and vacuum breakdown with electron-laser opposite beam // 7th International Conference on Multiphoton Processes (ICOMP VII), 1996, Garmisch-Partenkirchen, Germany. Book of Abstracts. P. B25.

\bibitem{Golovinsky1997} Golovinsky P.A. Electron-laser vacuum breakdown: cascade pair production in an intense laser field // Preprint VSU No. 97-3. Voronezh, 1997. 24 p.
\bibitem{ELI2018} ELI – Extreme Light Infrastructure Science and Technology with Ultra-Intense Lasers / ed. by G.A. Mourou, G. Korn. Bucharest: ELI-NP, 2018.

\bibitem{XCELS2015} XCELS – Exawatt Center for Extreme Light Studies: White Book / ed. by A.V. Kim et al. Nizhny Novgorod: IAP RAS, 2015.

\bibitem{XFEL2007} European XFEL Facility: Technical Design Report / ed. by M. Altarelli et al. Hamburg: DESY, 2007.

\bibitem{Bulanov2013} Bulanov S.V., Esirkepov T.Zh., Korzhimanov A.V. et al. QED cascades in intense laser fields // Physics-Uspekhi. 2013. Vol. 56. P. 429–455.

\bibitem{Narozhny2012} Narozhny N.B., Popov V.S. Production of electron-positron pairs in an intense laser field // Journal of Experimental and Theoretical Physics. 2012. Vol. 114. P. 1–32.

\bibitem{Sokolov2009} Sokolov I.V., Naumova N.M., Nees J.A., Mourou G.A. Pair creation in QED-strong fields // Physics of Plasmas. 2009. Vol. 16. P. 093115.

\bibitem{Golovinsky2018} Golovinsky P.A., Astapenko V.A., Mikhailov A.I. Elementary processes in electromagnetic field. Moscow: LENAND, 2018. 440 p. Chapter 5. P. 215–240.

\bibitem{Fedotov2010} Fedotov A.M., Narozhny N.B., Mourou G.A., Korn G. Limitations on the attainable intensity of high power lasers // Physical Review Letters. 2010. Vol. 105. P. 080402.

\bibitem{Heitler1954} Heitler W. The Quantum Theory of Radiation. 3rd ed. Oxford: Clarendon Press, 1954. 430 p.

\bibitem{SLAC1997} Burke D.L. et al. Positron production in multiphoton light-by-light scattering // Physical Review Letters. 1997. Vol. 79. P. 1626.

\bibitem{Bamber1999} Bamber C. et al. Studies of nonlinear QED in collisions of 46.6 GeV electrons with intense laser pulses // Physical Review D. 1999. Vol. 60. P. 092004.

\bibitem{LUX2012} Chen H. et al. Relativistic positron creation using ultra-intense short pulse lasers // Physical Review Letters. 2012. Vol. 108. P. 025004.

\bibitem{VULCAN2015} Sarri G. et al. Generation of neutral and high-density electron–positron pair plasmas in the laboratory // Nature Communications. 2015. Vol. 6. P. 6747.

\bibitem{Astra2019} Poder K. et al. Experimental signatures of the quantum nature of radiation reaction in the field of an ultraintense laser // Physical Review X. 2018. Vol. 8. P. 031004.

\bibitem{BELLA2023} Yakimenko V. et al. Experimental observation of radiation reaction in the collision of a high-intensity laser pulse with a laser-wakefield accelerated electron beam // Physical Review Letters. 2023. Vol. 130. P. 115001.

\end{thebibliography}
\end{document}